# On the use of decision tree regression for predicting vibration frequency response of handheld probes

Roberto San Millán-Castillo, Eduardo Morgado and Rebeca Goya-Esteban

*Abstract*— **This article focuses on the prediction of the vibration frequency response of handheld probes. A novel approach that involves machine learning and readily available data from probes was explored. Vibration probes are efficient and affordable devices that provide information about testing airborne sound insulation in building acoustics. However, fixing a probe to a vibrating surface downshifts sensor resonances and underestimates levels. Therefore, the calibration response of the sensor included in a probe differs from the frequency response of that same probe. Simulation techniques of complex mechanical systems may describe this issue, but they include hardly obtainable parameters, ultimately restricting the model. Thus, this study discusses an alternative method, which comprises different parts. Firstly, the vibration frequency responses of 85 probes were measured and labelled according to six features. Then, Linear Regression, Decision Tree Regression and Artificial Neural Networks algorithms were analysed. It was revealed that decision tree regression is the more appropriate technique for this data. The best decision tree models, in terms of scores and model structure, were fine-tuned. Eventually, the final suggested model employs only four out of the six original features. A trade-off solution that involved a simple structure, an interpretable model and accurate predictions was accomplished. It showed a maximum average deviation from test measurements ranging from 0.6 dB in low- frequency to 3 dB in high-frequency while remaining at a low computational load. This research developed an original and reliable prediction tool that provides the vibration frequency response of handheld probes.**

## I. Introduction

Airborne sound insulation is of central importance in building acoustics. Different international standards are well-known in scientific and practitioner fields, which involve estimating compliance with local government limits [1][2]. However, flanking sound transmission does not always form a part of these traditional standardised methods, which focus instead on sound pressure. This omission may hamper both the scope and correct implementation of optimal solutions [3][4]. Vibration velocity levels ($L_v$) on walls were used to collect more valuable and robust data about airborne and structure-borne insulation features. $L_v$ is the reference used in tests of flanking transmission in the laboratory as well as in-field measurements involving very recent ISO standards [4] and building acoustic simulation frameworks [5].

Moreover, $L_v$ is a classic variable used to define the radiated acoustic power of walls [6][7] and has proven useful in research conducted on building acoustics recently. Some relevant studies testify to the value of using vibration signals to estimate sound insulation: Discrete Calculation Method [8], the creation of sound and vibration handheld probes [9], the use of doppler laser to measure vibration signals [10][11][12], and the new

Transfer Path Analysis technique being applied to building acoustics. Those recent techniques also demonstrate the need for quite large numbers of accelerometer positions for measurements and the desirability of methods that obviate tedious and impractical testing techniques [13]. Still, while such techniques may be of great value, they are difficult to perform in real and common practitioner environments, due to equipment costs and the time that must be invested in collecting data. More straightforward methods that require wax or metal washers glued/cemented to the sensor [14] are not suitable when walls must remain aesthetically the same, without any flaws, after testing.

For the sake of simplicity and cost-effectiveness, an approach based on a general-purpose single-axis piezoelectric accelerometer with a metallic probe/rod is a good substitute. Such an approach prioritises comfortable, quick, affordable and less intrusive testing in practical cases. Airborne sound insulation estimation using vibration handheld probes is illustrated in *Fig.1*. However, signals from such a simple sensor can mislead $L_v$ assessment due to accelerometer resonance frequency shift-down and under-estimation of levels at specific frequencies. Other studies [15][16][17] presented details and further justification. Sensor manufacturers provide frequency response characteristics for fixings that are highly robust and stable (i.e. the screw-on method), but practitioners find it difficult to follow these procedures. A recent study in [18] proposed an innovative process for collecting the vibration frequency responses of handheld probes to estimate $L_v$, which also provided a wide-ranging database of probes. Vibration probes may be built in a simple, low-cost way using general-purpose piezoelectric accelerometers and available metal rods. To date, limited attention has been paid to handheld probes in vibration signal collection. Often thought to be restricted in use, handheld probes present valuable features as confirmed in [18].

Nevertheless, the information contained in a database might not always align with the probes required: different materials, different lengths or various sensors can vary depending on the working environment. Consequently, a prediction tool providing the vibration frequency response of a handheld probe based on its depicting parameters would be of great practical use. Simulation techniques of complex mechanical systems are available and well-defined [6], based on the general laws of dynamics. Still, parameters that are difficult to obtain are required, given internal material losses or the influence of hand movements. These parameters limit the generalisation capability of these models and may not be adequately adapted to situations in which some of that data is lacking. The research presented in this article confronted the challenge of acquiring the vibration frequency response of handheld probes using available and very easy to obtain input variables: sensor and probe specifications. This work differs from other studies based

The authors are with the Department of Signal Theory and Communications, Rey Juan Carlos University, Fuenlabrada 28943, Spain (e-mail: roberto.sanmillan@urjc.es).

on non-linear regression and highly detailed sensor data [19].

To the best of our knowledge, no study has specifically looked at the prediction of the vibration frequency response of handheld probes employing machine learning (*ML*) techniques. This research attempts to draw knowledge through *ML* and assesses its potential use for handheld probes. To implement and validate this prediction tool, the database generated in [18] was used. *ML* algorithms are a state-of-the-art technique in building acoustics. They provide prediction methods needing no prior physical information and using available features that are easy to compile. Recently, Artificial Neural Networks (*ANN*) have been widely used with remarkably good results for airborne sound insulation [20][21], sound absorption prediction [22], indoor noise [23] and hall acoustics estimation [24].

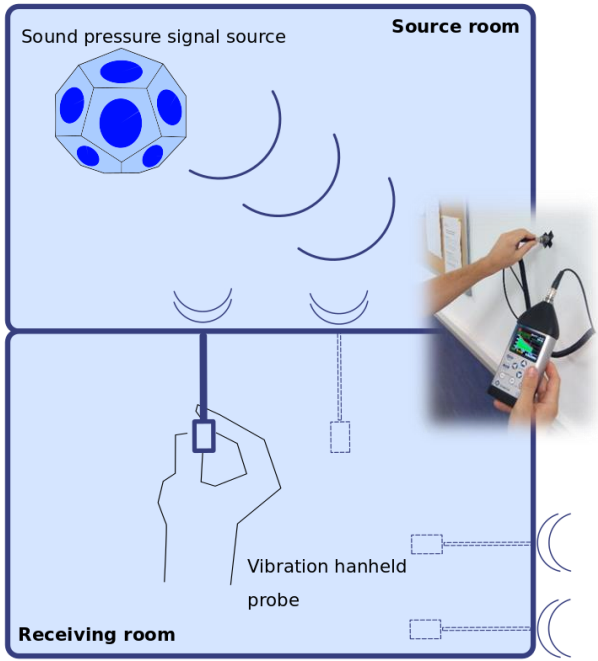

Fig.1. Conceptual figure regarding the targeted application: airborne sound insulation measurement with vibration signal collected by handheld probes. An acoustic pressure signal in the source room generates vibration on the walls of the receiving room. Time and spatial samplings of the vibration response are carried out on the walls of the receiving room by a handheld probe.

For the sake of interpretability, Linear Regression (*LR*) and Decision Tree Regression (*DTR*) were studied in this work. *LR* was initially used to tackle our problem as a classical and easy-to-interpret method. Then, *DTR* provides an alternative when non-linearities are present in our data and results need to be made interpretable. After that, *ANN* were explored as a more advanced method that also deals with non-linearities but implies some disadvantages when compared with the previous ones (e.g. *ANN* involve many hyperparameters, which need to be tuned; *ANN* offer no direct insights about the importance of the features in the input space). Furthermore, *LR* and *ANN* need feature scaling, but *DTR* methods do not require much effort while preparing data before their application.

To the best of the authors´ knowledge, *DTR* has never been applied to building acoustics tests or predicting the frequency response of sensors. However, *DTR* has been used in other fields with interesting results: the concentration estimation of gasses [25], air pollutants estimation [26], the prediction of radio urban propagation path loss [27], semiconductor yield loss analysis [28], even digital signal processing [29] and electricity energy consumption [30].

This research provides a novel proposal for the use of existing *ML* methods to predict the vibration frequency response of handheld probes in real-life measurement environments. The rest of the paper is organised as follows: *Section II* comprises materials and methods used for collecting data, together with the assessment and development of *ML* algorithms; *Section III* presents the results obtained from the predictions generated by the methods employed and an analysis of them; Finally, *Section IV* includes conclusions of this research.

## II.  MATERIALS & METHODS

### A.  Data Collection

*ML* methods require large amounts of data as input variables. Thus, data collection was the first step in our research. The target variable is the vibration frequency response of handheld probes. These responses were labelled for use within a supervised *ML* framework. A detailed description of all testing undertaken for data collection can be found in previous work [18]. All vibration measurements were taken in a controlled test rig. Sensor performance was assessed by the vibration acceleration level ($L_a$), referenced to $10^{-6}$ m/s$^2$. $L_a$ is preferred to $L_v$ if we compare sensor performance. Since this study focuses on building acoustics, the reference frequency range is from 50 Hz to 5 kHz, with a one-third octave band resolution as provided in most relevant standards [1].

An acoustic and vibration frequency real-time analyser was the data hub in experiments. A miniature bench-top electro-dynamic shaker generated a stable vibration. A rigid metal plate screwed to the shaker was the place where the measurements were taken. Photographs of different probe set-ups in the experimental platform are in *Fig.2*. These experiments simulate vibration signal collection in a real environment without the need for vibrating walls.

Five piezoelectric CCLD™ or ICP™ suitable sensors were used for testing (see *Table I*). To build probes, readily available metal rods were attached to the sensors by screwing them to their mounting holes. Two different general-purpose materials were used in these experiments: stainless steel and brass. The rigidity of materials is not generally available as a well-defined value, so the Young Modulus (*E*) was given by technical bibliography [7]: Brass *E* and steel *E* were 110 GPa and 210 GPa respectively. Furthermore, the direct handheld sensor used by the practitioner without any probe (*Plain,* see *Fig.2-b*) was modelled with much higher rigidity compared to the probes, *E* thus becoming 210.000 GPa.

This study tested a range of eight nominal lengths of probes for each pair of material and sensor. Therefore, 80 different probe combinations were analysed, along with 5 *Plain* cases. *Table II* presents all precise specifications of probes, which combined with sensors in *Table I*. In total, 85 frequency response measurements from different vibration probe set-ups were collected. As is common in building acoustics, the frequency response is obtained from 5 s block average time measurements [1][4]. Every single measurement was 17 s long. Firstly, two seconds were discarded due to high variability resulting from the initial instabilities of probe location. Secondly, each measurement file was split into three 5 s files to obtain more data within a typical time measurement duration. All measurement types were repeated eleven consecutive times by the same operator in order to simulate real-measurement changes due to the probe location.

<div align="center">

TABLE I
ACCELEROMETER TYPES INVOLVED IN EXPERIMENTS
</div>

| Sensor Sensitivity [mV/g] | Sensor Weight [gr] | Sensor connector location | Sensor Type / Manufacturer |
|---|---|---|---|
| 10.03 | 5.9 | Top-mounted | 352C04 / PCB[a] |
| 46.69 | 8.6 | Side-mounted | 4533-B-004 / B&K[b] |
| 101.40 | 5.8 | Side-mounted | 352C33 / PCB |
| 526.70 | 8.6 | Side-mounted | 4533-B-002 / B&K |
| 975.00 | 25 | Top-mounted | 352B / PCB |

a, b PCB and B&K are two accelerometer manufacturers: PCB Piezotronics Inc. and Brüel&Kjaer Sound & Vibration measurements A/S, respectively.

<div align="center">

TABLE II
PROBE TYPES INVOLVED IN EXPERIMENTS
</div>

| Probe Length [cm] | Probe weight [gr] | Probe Material [E] |
|---|---|---|
| 6.4 | 7 | Brass |
| 12.4 | 13 | Brass |
| 19.9 | 25 | Brass |
| 25 | 28 | Brass |
| 29.7 | 39 | Brass |
| 40 | 52 | Brass |
| 49.9 | 53 | Brass |
| 101 | 135 | Brass |
| 6.2 | 5 | Steel |
| 12.5 | 12 | Steel |
| 19.8 | 23 | Steel |
| 24.5 | 24 | Steel |
| 29.6 | 36 | Steel |
| 39.9 | 45 | Steel |
| 49.6 | 48 | Steel |
| 100.5 | 119 | Steel |

### B. Performance Assessment of Models

Data labelling is a crucial task for supervised *ML* estimators. For the sake of affordability, the data features of this work are simple and easily available. Hence, the features we chose for our proposed *ML* algorithms are namely: *sensor sensitivity* in mV/g, *sensor weight* in gr, *sensor connector location*, *probe length* in cm, *probe material* in GPa and *probe weight* in gr. Every vibration frequency response of the handheld probes was identified and labelled by their six features so that the algorithm learned from the measurements provided.

The preferred performance characteristic of predictors is its generalisation capability. With this in mind, the *train* data set included most of the measurement samples to train the model. *Validation* data comprised data that is not included in the *train* data set. This procedure allowed us to evaluate the trained model with measurements that differed from those employed in the *train* data set. The selection of samples for *train* and *validation* data sets was random and performed in the cross-validation stage (*CV*).

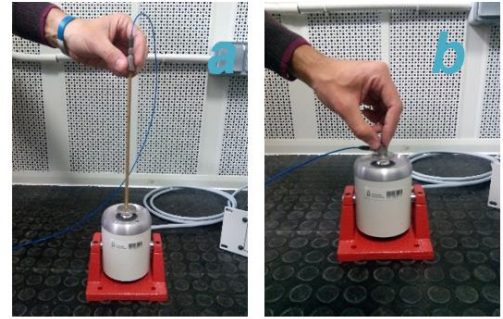

Fig.2. Different testing set-ups in data collection stage: a) 50 mm long brass probe on a top-mounted connector accelerometer; b) Direct handheld (Plain) set-up of a side-mounted connector accelerometer [18]

*CV* helps evaluate a model made from a small data set in which no samples should be wasted, such as the one in this work. Many studies assess models using *CV* in train/test data sets; however, in the present work, a more robust evaluation process was applied using *CV* in train/validation data sets and a final evaluation in a hold-out test set. *K-fold CV* was the method employed in this study, setting-up five-folds as widely recommended in the literature [31]. Thus, the *train* data set split into *k* smaller sets, and the model was trained using *k-1* folds. The resulting model is validated in the remaining data fold. Finally, the performance of the model was assessed using the *k-fold* average score. When *validation* scores were successful, we proceeded to the *test* stage.

A real situation was emulated to evaluate the model as follows: from the 85 measured probes, all measurements from three probes, the *test* data set, were excluded from the *validation* and *train* data sets. Hence, the *ML* regressors were assessed with samples never checked by the algorithms, namely the *test* data set. After splitting data, 2706 blocks of measurements were available for *train* and *validation* data sets (77.2% for training, 19.3% for validation, from all data), and 99 for *test* data set (3.5%). Eventually, a model capable of a reasonable generalisation might be expected for new samples. The three *test* probes evaluated were a selection of representative cases among the 85 probes types. *Table III* describes the six features of those three test probes.

<div align="center">

TABLE III
FEATURES OF TEST PROBES
</div>

| Sensor Sensitivity [mV/g] | Sensor Weight [gr] | Sensor connector location | Probe Length [cm] | Probe weight [gr] | Probe Material [E] |
|---|---|---|---|---|---|
| 10.03 | 5.9 | Top-mounted | 6.4 | 7 | 110 (Brass) |
| 101.40 | 5.8 | Side-mounted | 39.9 | 48 | 210 (Steel) |
| 526.70 | 8.6 | Side-mounted | 24.5 | 23 | 210 (Steel) |

The present work researched the potential of *ML* at predicting probe frequency response. A *ML* evaluation method is performed by scoring regression performance versus labelled data. Accordingly, the goal was to assess the error of our model regarding measured data through comparisons. *Mean squared error (MSE)* is the intuitive score employed in this study to evaluate the error. Since *MSE* calculates an average and some subtractions, dB values were turned into linear values to carry out model training and scoring correctly, as defined in (1), (2) and (3):

$$a_{rms,f} = a_{ref} 10^{L_{a,f}/20} \ [m/s^2], \tag{1}$$

$$\widehat{a_{rms,f}} = a_{ref} 10^{\widehat{L_{a,f}}/20} \ [m/s^2], \tag{2}$$

$$MSE(a_{rms,f}, \widehat{a_{rms,f}}) = \frac{1}{n_s} \sum_{i=1}^{n_s} (a_{rms,f,i} - \widehat{a_{rms,f,i}})^2, \tag{3}$$

where $a_{rms,f}$ and $\widehat{a_{rms,f}}$ are the root mean squared values of the vibration acceleration of the measured data and predicted results respectively; $n_s$ is the number of samples involved in *MSE*, indexed by $i=\{1...n_s\}$; and $f$ refers to the frequency bands index from 50 Hz to 5 kHz. $L_{a,f}$ and $\widehat{L_{a,f}}$ are the corresponding vibration acceleration levels in dB referenced to $a_{ref} = 10^{-6}$ m/s². However, for the sake of interpretability in daily practitioner contexts, results are also shown in terms of the deviation from the real measured data. Hence, differences between $\widehat{L_{a,f}}$ and measured data can be assessed. All the cases included in the *test* data set considered the average deviation for every one-third octave band. Equations (4) and (5) define this new parameter.

$$y_{dif,f} = |\widehat{L_{a,f}} - L_{a,f}| \ [dB], \tag{4}$$

$$DevdB(L_{a,f}, \widehat{L_{a,f}}) = 10 log(\frac{1}{n_c} \sum_{m=1}^{n_c} 10^{\ y_{dif,f,m}/10})[dB], \tag{5}$$

where $n_c$ is the number of testing cases, which are 99, indexed by $m=\{1...n_c\}$ and $f$ is the frequency bands index from 50 Hz to 5 kHz. $\widehat{L_{a,f}}$ is the dependent variable prediction of the model based on the features used and $L_{a,f}$ is the measured data, for each frequency band ($f$) and each case ($m$) in the *test* data.

$L_a$ values from our dataset appeared remarkably different within the frequency range, see *Fig.3*. Thus, every model proposal contains 21 predictors corresponding to the 21 one-third frequency bands of interest. Throughout this study, the *ML* strategies applied to each frequency band were the same.

The implementation and evaluation of the model were performed by algorithms originally developed by the authors, with Python 2.7 programming language [32]: using a scientific computing library, *NumPy* [33]; a data analysis library, *Pandas* [34]; *Matplotlib* library [35] assisted in the plotting of results. Specific functions involving *ML* were based on a machine-learning library, *Scikit-learn* [36], essentially: *LinearRegression*, *DecisionTreeRegression*, *MLPRegressor* and *GridSearchCV*. Algorithms were run in an i5™ processor, 8GB RAM 64bits windows laptop.

## C. Methods

*LR*, *DTR* and *ANN* were the *ML* methods explored in this study.

### C.1 Linear Regression

*LR* was based on a classical least-squares method [31]. The linear model defines $\widehat{L_{a,f}}$ as a linear combination of the six original features, in (6):

$$\widehat{a_{rms,f}} = \ \widehat{\beta_0} + \widehat{\beta_1} X_1 + \widehat{\beta_2} X_2 + ... + \widehat{\beta_6} X_6 + \varepsilon \ [dB], \tag{6}$$

where $\widehat{a_{rms,f}}$ is the outcome variable or prediction, $\beta_0$ is the intercept, $\{\widehat{\beta_1}...\widehat{\beta_6}\}$ are the estimated coefficients for each feature that minimise the residual sum of squares between measurements and linear predictions, $\{X_1...X_6\}$ are the independent variables or features, and $\varepsilon$ represents the errors. The input features were standardised to zero mean and unit variance.

### C.2 Decision Tree Regression

*DTR* is a non-parametric supervised learning method that predicts the value of a target variable through simple decision rules inferred from measured training data and their features [31]. For the sake of the interpretability of the model, *DTR* was considered the next step after *LR* toward approaching this research problem. *DTR* is simple to interpret since it highlights the importance of features and provides visible results on the model as a tree structure. Conversely, *DTR* might also create overly complicated decision rules that may lead to overfitting and may impair the generalisation of models. However, hyperparameter tuning can control these drawbacks. The algorithm performed in this study is CART [31][37], which operated with the following regression criteria: the input feature space is divided into $M$ non-overlapping regions $R_m$, $m=\{1...M\}$. For every observation that falls into the region $R_m$, the predicted $\widehat{a_{rms,f,i}}$ is the mean of the response values for the training observations in $R_m$. The goal is to find the regions that minimise *MSE (Rm)*, which grow the regression tree, given by (7).

$$MSE(R_m) = \frac{1}{n_{samples}} \sum_{m=1}^{M} \sum_{i \in Rm} (a_{rms,f,i} - \widehat{a_{rms,f,i}})^2, \tag{7}$$

To this end, an iterative greedy algorithm is used. Firstly, it finds the input feature $X_m$ and cut-point $s$ such, that splitting the feature space into regions $\{X|X_m < s\}$ and $\{X|X_m \geq s\}$ leads to the highest possible reduction in *MSE(Rm)*. Next, the process is repeated but splitting one of the two previously identified regions. The process continues until a stop criterion is reached.

This study considered two hyperparameters values for *DTR* tuning. The first hyperparameter was the number of features considered when the best split was chosen (*MaxFeat*). Regarding *MaxFeat*, the following facts were borne in mind: the best-selected division considered all of the features and the minimum number of samples required to split a node was two. After that, the minimum number of samples needed to be at a leaf node was one, and there was no limit to the number of leaf

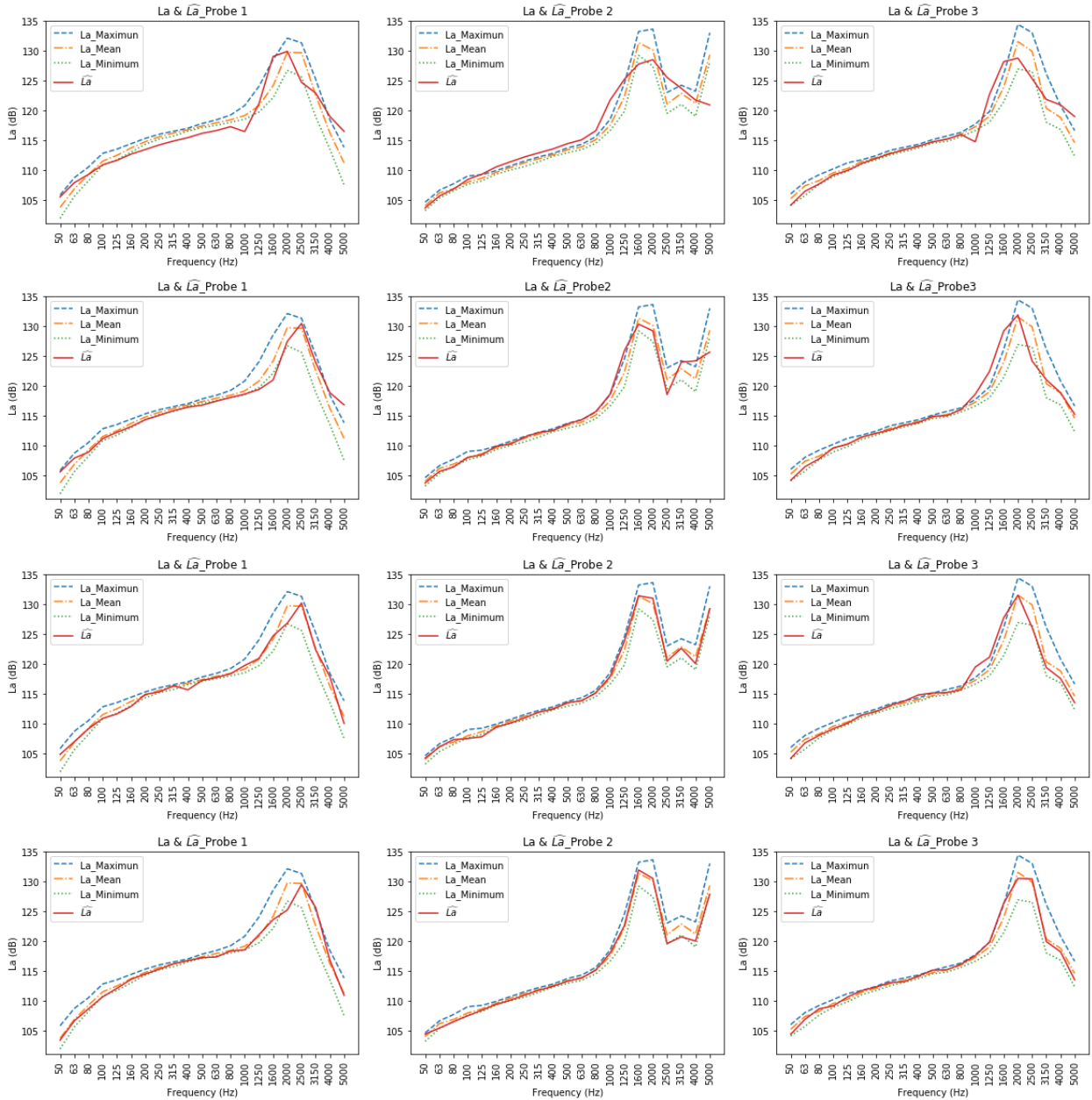

Fig.3. Matching of the predictions of four representative models regarding measured data, through a comparison of aggregated values of measurements ($L_a$: maximum, minimum and mean values) and model prediction, $\widehat{L_a}$: first row, *LR* ; second row, *ANN*; third row, *DTR (2xGrid)*; fourth row, *DTR (2xGrid+4F)*. The three different *test* probes were considered for each column.

nodes. The second hyperparameter was the maximum number of splits for a sample (*MaxDepth*).

Regarding *DTR* model selection, two approaches were undertaken for tuning *DTR* hyperparameters. The first attempt only took into account a *MaxDepth* value of 14, to prevent a heavy computational load. The second procedure exhaustively worked two hyperparameters of the estimator, namely: *MaxDepth*, as the previous tuning strategy; and the *MaxFeat*, considering the six original suggested features of the model. Hence, a grid of values of the chosen hyperparameter/s was generated, which comprises all possible combinations. That parameter space contained all candidates to be the best model. The performance of the model was assessed by *MSE*. The *CV* was performed at this stage, as described in *Section II.B*. All candidates were arranged according to average validation *MSE* scores of *CV* folds. Finally, the best candidate was the selected model for further evaluation with the three *test* probes.

### C.3  Artificial Neural Networks

*ANN* extract linear combinations of the inputs as derived features and then model the target problem as a non-linear function of those features. This advanced method works with a number of hyperparameters, which are not very intuitive in general. Moreover, *ANN* tuning optimisation problems can also lead to overfitting when regularisation is discarded. Some authors even stated that the tuning of *ANN* is close to its an art [31].

Typically, the structure of the *ANN* comprise an input layer that includes a set of neurons representing the model input features {$X1 ... X6$}, followed by one or more hidden layers that transform the values of the input layer using a non-linear activation function. Each hidden layer has a number of neurons selected as a design parameter, and each neuron connects to the neurons on the previous and the following layers by a weighted link. The weights (*W*) are estimated in the *ANN* training stage. Finally, the output layer operates the previous values and provides the estimated regression value as output ($\widehat{a_{rms,f}}$). The *ANN* algorithm employed in this study was based on a single hidden layer back-propagation network since it is one of the most widely used due to its simplicity and efficiency. This model optimises the squared-loss (*R*) updating *W* in (8):

$$R(\widehat{a_{rms,f}}, a_{rms,f}, W) = \frac{1}{2} \left\| \widehat{a_{rms,f}} - a_{rms,f} \right\|_2^2 + \frac{\alpha}{2} \|W\|_2^2, \quad (8)$$

where $\widehat{a_{rms,f}}$ is the prediction, where $a_{rms,f}$ is the measured data, and $\frac{\alpha}{2} \|W\|_2^2$ is a term that penalises complex models.

*ANN* are sensitive to feature scaling. The studied data set presents substantial differences among features ranges, and a pre-processing of features data was carried out, namely: the removal of mean and scaling to unit variance. For the sake of simplicity, *ANN* model selection was conducted thoroughly considering some of the more relevant hyperparameters. This research took into account a specific range of five hyperparameters; more details of them are available in [36]. The number of neurons in the hidden layer ranged from 1 to 20 units. The different activation functions of the hidden layer were the *Logistic Sigmoid*, the *Hyperbolic Tangent* function, the *Rectified Linear Unit*, and the *Identity* functions. Then, the weight optimisation approaches were *L-BFGS (quasi-Newton method)*, *Stochastic Gradient Descent*, and *Adam Stochastic Gradient*. Finally, the learning rate strategies considered were the *Constant*, *Inverse Scaling Exponent*, and *Adaptative* rates. The maximum number of iterations until convergence was 200. As in the *DTR* model evaluation, this procedure resulted in a grid of all possible candidates to be the best model for *ANN*. After assessing the *MSE* and performing the *CV*, the best candidate model was selected.

## III. RESULTS AND DISCUSSION

The strategy developed to identify our best final model was as follows. Firstly, a simple approach to the problem was carried out with *LR*, but the results were not entirely successful. Secondly, *DTR* was studied scrutinising only one hyperparameter, which delivered more promising results compared to *LR*. After that, two hyperparameters fine-tuned the *DTR* model. Then, a simple *ANN* approach was employed as a method already used in building acoustics whose results improved *LR* ones but remained worse than *DTR* models. The goal was to draw the best model at every stage, considering the best scores and the most simplified model structure. Once the chosen model was found, another fine-tuning step was carried out for *DTR*, based on feature selection. Model evaluation was based on *MSE* and *DevdB* values.

### A. Starting Approaches

This section addresses the issue with simple *LR*, *DTR* and *ANN* models and explores their performances. The first approach was *LR*, employing the six features taken into consideration. *Fig.3* shows the performance of the predictions of different *ML* methods for the 3 *test* probes with respect to their vibration frequency responses. $\widehat{L_a}$ represents model prediction whereas $L_{a\_Maximum}$, $L_{a\_Minimum}$, and $L_{a\_Mean}$ represent the maximum, minimum and mean values of $L_a$ for the *test* probes, respectively.

The first row in *Fig.3* refers to *LR* and reveals some differences between the predictions and the measured values. Thus, there is a remarkable disagreement in the resonance area, in high- frequency: inaccurate resonance prediction in terms of frequency band and some differences in resonance boosting and after-resonance attenuations. Besides, predictions were not precise broadband since they hardly remain within the measurement limits. However, *Fig.3* describes a reasonable estimation in low-frequency for all test probes.

An additional comparative analysis among *DevdB* of *LR,* several *DTR* models, and *ANN* is presented in *Fig.4*. For a straightforward comparison, *Table IV* includes the *DevdB* values of some of the explored models. *DevdB* provides an aggregated metric that accounts for all the test data set and simplifies results. *DevdB* analysis indicates that a better approach than *LR* might be worthy for frequencies bands beyond 1 kHz since the *DevdBs* of *LR* were the highest ones. *LR* was revealed to be a simple model that yields fair predictions in the low-frequency range, but in higher frequency bands the deviations were considered too high. Therefore, it was decided to experiment with some other *ML* algorithms.

An interesting alternative to *LR* is *DTR*, as explained in *Section II*. The next approach considers six features and one of the critical hyperparameters of *DTR* in terms of interpretability: *MaxDepth*, in which the range of study was from 1 to 14. *DTR* is scale-invariant, and data distribution needs no assumptions. Hence, pre-processing data stage is simpler with *DTR* than with *LR* and *ANN*. The new addressed models are: an immediate *DTR* with *MaxDepth* = 14 for all frequency bands, denoted as *DTR (14)*; and a more thorough *DTR* was carried out involving the search for the best *MaxDepth* at each frequency band, denoted as *DTR (1xGrid)*.

TABLE IV
DevdB FOR DIFFERENT MODELS

| MODEL | DevdB |
|---|---|
| *LR* | 38.6 |
| *ANN* | 27.4 |
| *DTR (14)* | 18.7 |
| *DTR (1xGrid)* | 19.9 |
| *DTR (2xGrid)* | 19.7 |
| *DTR (2xGrid+4F)* | 17.5 |

The results are promising, as can be seen in *Fig.4* and *Table IV. DTR* models provided lower *DevdB* than *LR*, and eventually than *ANN*. Decreasing *DevdB* leads to less uncertainty for probe predictions and therefore, a more reliable model. Regarding *DevdB*, both *DTR* models considered at this step showed slight differences. The main contribution of *DevdB* lays on the resonance area for all models, from the 1 kHz band on. Regression models predicted rightly in the low-frequency range and remarkable differences raised in the resonance area. *Fig.3* and *Fig.4* show these issues.

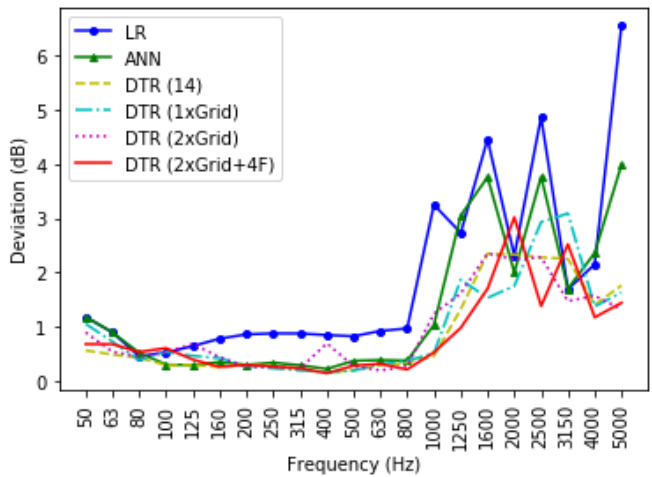

Fig.4. Comparison of *DevdB* of the 3 *test* probes for *LR, ANN, DTR (14), DTR (1xGrid), DTR (2xGrid), and DTR (2xGrid+4F)* for all frequency bands of interest.

However, our first *DTR* attempt, *DTR (14)*, provides the most cumbersome model since it works with all features and the maximum depth for all frequency bands. On the other hand, *DTR (1xGrid)* improved the previous model by decreasing *MaxDepth* at specific frequency bands and, consequently, the computational load. In sum, *DTR (1xGrid)* presented a more straightforward structure, and thus, a more interpretable model than *DTR (14)*. *Fig.5* offers information about complexity in the *DTR* models. Each marker is defined by a different shape that represents a different *DTR* model. One marker is included for each frequency band and model; therefore, a higher colour intensity means a larger number of overlapped markers. Markers near the upper right corner indicate more complex models than markers near the lower-left corner.

Once the previous *LR* and *DTR* experiments finished, *ANN* were assessed. The second row of *Fig.3* presents *ANN* predictions. All the features were considered along with the range of hyperparameters detailed in *Section II.C.2* This non-linear approach outperformed *LR* values remarkably up to 1kHz-1.25kHz one-third octaves. In the resonance area, *ANN* resulted in good predictions, but some level differences seemed to appear with respect to boosting and attenuations. *Table IV* and *Fig.4* indicate these disagreements. Generally speaking, there was a remarkable difference between the low- frequency range and resonance area predictions. In low-frequency, the use of one to six neurons in the hidden layer, the *L-BFGS* solver, the *Logistic Sigmoid* activation function, and a *Constant* learning rate were the typical tunings. However, in the resonance area, *ANN* accurately predicted when employing *Adam Stochastic* solver, the *Hyperbolic Tangent* activation function, an *Adaptative* learning rate, and the use of up to 19 neurons in the hidden layer. Although the *ANN* approach may present high expressive power to draw knowledge from data, it resulted in some lower interpretable models than *LR* or *DTR. ANN* also offer higher computational load, in part due to the hyperparameters tuning. Hence, *ANN* required from five to seven days to obtain outcomes while *DTR* or *LR* took between five and ten minutes.

The analysis of the previous results led us to focus on the more balanced technique for this data: *DTR*.

### B. Final Approach

The results discussed in *Section III. A* were accurate with regard to the vibration response of probes since precision sensors usually provide laboratory calibration certificates with a tolerance of +/-3 dB. The use of *DTR* corrected the unbalanced results at different frequency ranges of *LR* and *ANN* with remarkable success, both in *DevdB* and model structure terms. This section is devoted to investigating whether the *DTR* performance might be enhanced.

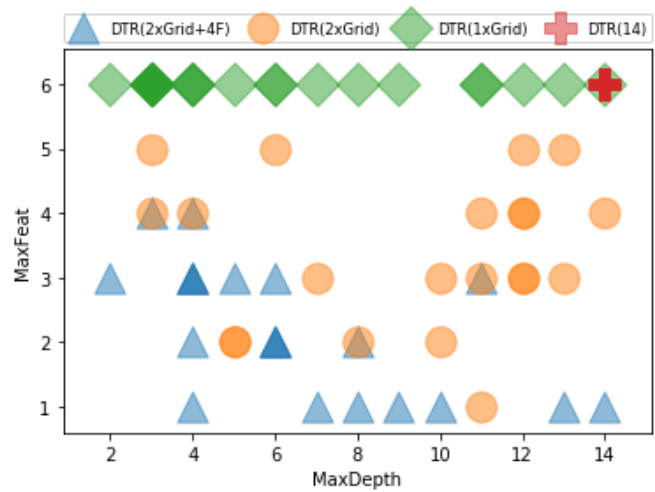

Fig.5. Scatter plot showing different *DTR* model structures in terms of *MaxDepth* and *MaxFeat* of all frequency bands. The models compared in this figure are *DTR (14), DTR (1xGrid), DTR (2xGrid)* and *DTR (2xGrid+4F)*.

Both *MaxDepth* and *MaxFeat* were involved simultaneously in tuning the model, denoted as *DTR (2xGrid). Fig.4* and *Table IV* confirmed that the predictions agreed to those provided by the previous *DTR* algorithms. *DTR (2xGrid)* slightly decreased the maximum *DevdB* at high frequencies regarding *DTR (1xGrid)* but providing a less complex model, see *Table IV, Fig.4* and *Fig.5.* Thus, the study continued using this *DTR (2xGrid)* model. The individual performance of the *DTR (2xGrid)* model showed a maximum prediction difference of 0.8dB at low frequencies and up to 2.3 dB in the resonance area. Also, *Fig.3* confirmed that *DTR (2xGrid)* provided better predictions than the *LR* and *ANN* models did. *DTR (2xGrid)* agreed remarkably at low frequencies and resonance frequency bands were estimated correctly, even with a second resonance in the response (*probe 2*). Resonance boosting/attenuation also was quite accurate within measurement limits in most cases. These considerations showed that a tuning carried out with two hyperparameters is a worthy refinement for the *DTR* model.

### B.1   Feature Selection

Once the two chosen hyperparameters optimised the *DTR* model, this section is devoted to examining the model regarding its features. An analysis of the model performance was conducted concerning the importance of features. A reduction of features would benefit the model complexity. Therefore, it was wise to weight the importance of each of the independent variables. To this end, this research might obtain a trade-off between model performance and the least number of features.

The procedure was initiated working out the relative importance of the six original features, using a Gini impurity index criterion [31]. Next, the least important feature was discarded and a new *DTR (2xGrid)* was tuned with *MaxDepth* and *MaxFeat* using only the five remaining features. This methodology was repeated with five, four, three and two features. Finally, a one-feature model was also considered. These models are denoted *DTR (2xGrid+nF)*, where *n* is the number of features included in the *DTR (2xGrid)* model. Note that this procedure considers possible non-linear interactions among features. A new *DTR* model was obtained every time a feature was discarded, and the ranking of feature importance updates for the remaining features. Finally, features were discarded in the following order: *sensor connector location*, *sensor weight*, *sensor sensitivity*, *probe material*, *probe weight,* and *probe length*.

At this point in the development of the model, it was essential to analyse the performance of the *training* and *validation* data sets. *Fig.6* and *Table V* describe the *DTR (2xGrid)* evaluation of the model and its generalisation in terms of the number of features considered. The performance was assessed with the sum of *MSE* in all one-third frequency bands (*MSE sum*) in *train* and *validation* data sets. Regarding the *MSE Training sum*, it presented a typical behaviour. Training scores were better as more features were involved in the model. However, the generalisation of models depends on validation scores rather than on training scores. *Validation* scores of *DTR (2xGrid+4F)* revealed that beyond four features is not worth adding more features since *MSE Validation sum* remains nearly constant. *Table V* also remarks that *DTR (2xGrid)* provides better validation scores than all the previous analysed models. *DTR(2xGrid+4F)* also outperformed *LR*, remarkably, and *ANN* concerning *MSE Training* and *Validation sums*, as shown in *Table V*.

A physical interpretation of the vibration frequency responses might be consistent with results obtained by *ML* feature selection and the database of measurements used in [18]. *Sensor connector location* and *weight* slightly influenced responses in some *Plain* probes as well as in the heaviest sensor, but that effect disappears in the other probes. *Sensor sensitivity* apparently played a secondary role in the model since all accelerometers involved were precise enough for the circumstances of the experiment. Unexpectedly, the predictive models obtained more valuable information from *Sensor sensitivity* than from other features. *Probe Material, weight*, and *length* were the most relevant features since they caused real changes in the vibration frequency response of the probes. They are responsible for the sensor resonance downshift and the rise of new resonances due to modifications to rigidity caused by the sensor fixing technique. *Probe length* was the most important feature, followed by *probe material*. As for boost or attenuation in response, *probe weight*, and *probe material* were critical.

TABLE V
MSE VALIDATION & TRAINING SUMS FOR DIFFERENT MODELS

| MODEL | MSE Training sum | MSE Validation sum |
|---|---|---|
| *LR* | 6.2 | 24.7 |
| *ANN* | 3.7 | 4.6 |
| *DTR (14)* | 1.3 | 4.4 |
| *DTR (1xGrid)* | 2.2 | 3.8 |
| *DTR (2xGrid)* | 1.3 | 3.4 |
| *DTR (2xGrid+4F)* | 2.1 | 3.4 |

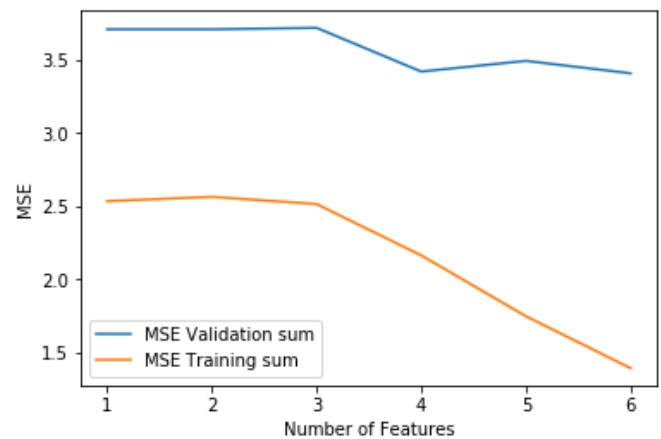

Fig.6. *Comparison of MSE Validation sum and MSE training sum regarding the number of features selected in the DTR (2xGrid) model.*

This research deals with a suitable range of probes for building acoustics, meaning that all alternatives are effective to a large extent; there are no bizarre or unusual solutions or probe parts. Thus, *MSE Validation sum* results are similar despite the features included in the model. Nevertheless, the suggested feature selection procedure provides fine-tuning of the model in generalisation terms according to scores shown in *Fig.6* and *Table V*. Therefore, *DTR (2xGrid+4F)* seems to be the best choice concerning *MSE Validation sum*.

An examination of the structure issues of *DTR (2xGrid+4F)* needs to be addressed. In terms of simplifying the model structure, *Fig.5* clearly shows a significant drop of *MaxDepth* for *DTR (2xGrid+4F)*. *MaxFeat* improved since only four features were employed, fewer than the rest of the evaluated models in *Fig.5*. Not only an advantage in prediction results was found, but also added value concerning model complexity.

In sum, this research identified *DTR (2xGrid+4F)* as the least complex model that accomplished the best generalisation score.

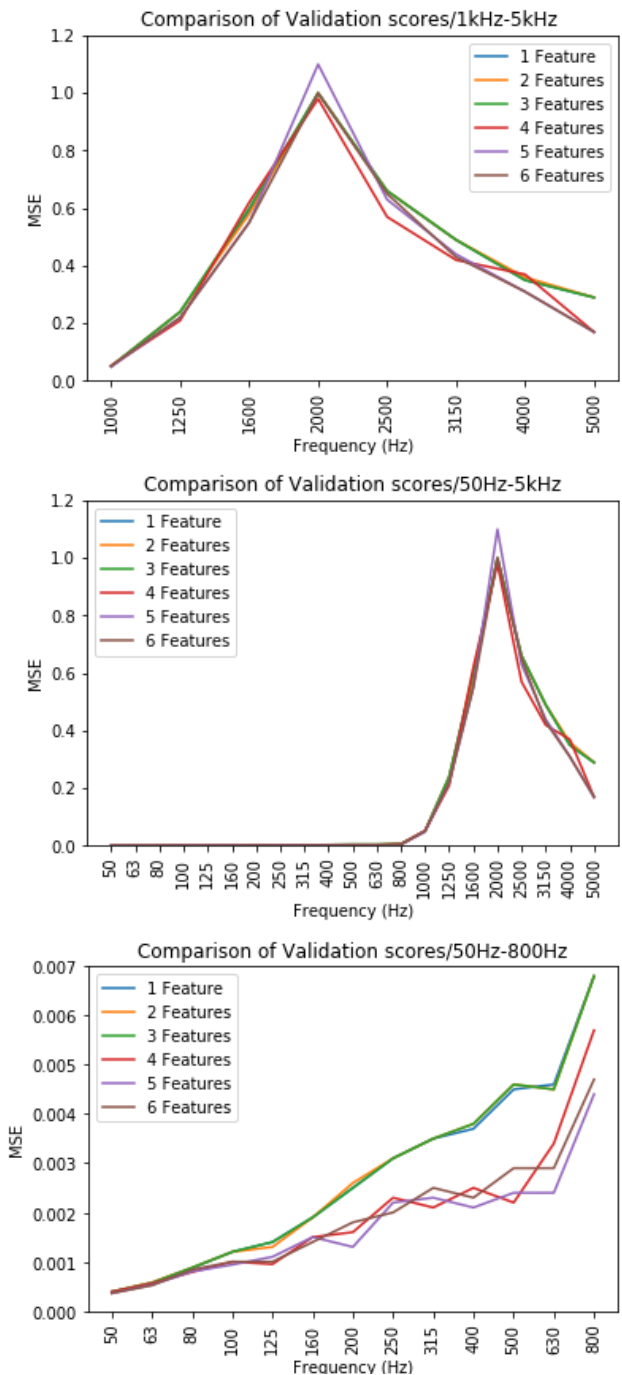

Fig.7. Comparison of *MSE Validation sum* regarding the number of features in the *DTR (2xGrid)* model, from one feature to six features. The upper panel is a general view of all bands of interests. The two lower panels focus closely on the general *probe useful bandwidth* and the resonance area, respectively.

### B.2 Analysis in Frequency and Uncertainty Terms

A detailed analysis was explored in the frequency domain since we confirmed substantial differences between prediction accuracy in the low and high-frequency ranges. *MSE Validation sum* analysis needs to be more deeply scrutinise since it sums all frequency validation scores and overlooked differences among frequency bands. *Fig.7* provides a comparison of the *MSE Validation sum* of *DTR (2xGrid)* models versus frequency in, ranging from 1 to six features. Predictions of the high-frequency range remarkably influenced the *MSE Validation sum*.

From 1 kHz, predictions became difficult due to resonances and its random variability in terms of vibration frequency response. Hence, *MSE* values in high-frequency bands were much larger than in lower ranges, see *Fig.7* lower panels. Consequently, they are the main contribution to the *MSE Validation sum*.

However, predictions did agree accurately within the *probe useful bandwidth* defined in [18]. The *probe useful bandwidth* indicates the frequency range where the studied probe presents a frequency response within typical tolerance values of laboratory calibration of accelerometers, which are widely recognised as the more accurate ones. Depending on the probe, it ranges from 50 Hz to 800-1.6 kHz. The *Probe useful bandwidth* shows where the probe frequency response is nearly the same as the accelerometer response and is interchangeable.

The number of features also influenced the performance of the model. In *Fig. 7*, results up to 3 features almost overlapped. The consideration of more features followed some improved results in low-frequency predictions. However, from four features on, this was insignificant with respect to *MSE*. The four-features model remained the most valuable one in terms of critical resonance bands between 1.6 kHz to 3.15 kHz. In the highest frequency range, the improved performance of *DTR (2xGrid+4F)* is remarkable.

TABLE VI
VALIDATION STANDARD DEVIATION SUM IN FREQUENCY

| **MODEL** | *Standard sum* | *Δ Standard sum [%]* |
|---|---|---|
| *DTR (2xGrid)* | 1.41 | 0 |
| *DTR (2xGrid+5F)* | 1.60 | 13 |
| *DTR (2xGrid+4F)* | 1.49 | 5 |
| *DTR (2xGrid+3F)* | 1.84 | 30 |
| *DTR (2xGrid+2F)* | 1.88 | 33 |
| *DTR (2xGrid+1F)* | 1.88 | 33 |

The variability of a regression model provides information about model uncertainty. *DTR* is a statistical method that offers different results depending on the samples used in the *train* and *CV* stages. Uncertainty was assessed employing the standard deviation of the scores obtained in the *CV* stage. *Table VI* details the sum of standard deviations in all the frequency bands (*Standard sum*) of the *DTR (2xGrid)* models. The six-features model was very similar to the four-features model according to scores. However, *DTR (2xGrid+4F)* was more uncertain in terms of *Standard sum*. The comparison of uncertainty is presented in terms of the increase of *Standard sum* (*Δ Standard sum*) of each model referenced to *DTR (2xGrid)*, see *Table VI*. *DTR (2xGrid+4F)* remained close, 5% in terms of *Δ Standard sum*, whereas the one-feature model disagreed 33%. Such deviations permit a practitioner to gain knowledge about the reliability of the model. Hence, *Standard sum* values provide information to select the most suitable model according to the practical situation concerned. Models involving less than four features showed worse results in terms of *Standard sum* and therefore, more uncertainty regarding the accuracy of predictions.

## IV. CONCLUSION

This research explores a novel method for predicting the vibration frequency response of handheld probes based on existing ML techniques. For the first time, algorithms as *LR*, *DTR*, and *ANN* were used in such a problem.

Model features are readily available specifications of sensors and probes. Thus, this study discards parameters that are generally difficult to obtain and are included in complex theoretical models. The results of this research indicate the reliability of the proposed prediction tool. To date, the vibration frequency response of handheld probes has been scarcely available. Using this research, both scientists and practitioners have at their disposal a prediction framework that appeared successful for the tested probes.

Thus, handheld probes could be designed for certain *ad hoc* circumstances and know their vibration frequency responses before collecting data. This prediction model enhances the collection of vibration signals avoiding errors coming from weightings in the frequency response of sensors due to the probe features. Besides, these predictions would help to optimise the *probe useful bandwidth*. Accordingly, measurements could be nearly equivalent to those collected from more robust fixing methods of accelerometers but improving operation facts, due to the advantages of a handheld device. This research is a useful contribution to the field of building acoustics that might extend its applicability to other disciplines related to vibrations, within a similar frequency and dynamic ranges.

*LR* and *DTR* were the methods evaluated for the sake of interpretability. The application of both techniques in our regression problem concludes that the use of *DTR* remarkably offers better results than the use of *LR*. However, this difference in performance could not be predicted since it depends on the nature of the problem. All performance metrics of the models and developing details are available for the research community. Thus, the more suitable model may be selected depending on different priorities.

*ANN* models were also analysed to verify its potential on this dataset since they are a reference *ML* method employed in building acoustics. The experiments provided better results than *LR* and worse than *DTR* in terms of *validation*, *training* scores, and *DevdB*. A simple but thorough approach of *ANN* was applied. Despite this fact, there was also a thousandfold increase in computational time regarding *DTR*, which add another disadvantage to *ANN* for this dataset. To keep a simple approach, a further and more advanced setup analysis remained out of the scope of this research.

The final choice model for our dataset was a *DTR* optimised by two hyperparameters (*MaxFeat* and *MaxDepth*) and fine-tuned with only four of the six original features. *Probe length*, *probe weight*, *probe material*, and *sensor sensitivity* appeared as the most valuable features for modelling frequency response. *DTR (2xGrid+4F)* was the best model regarding *DevdB* but not the best with respect to *MSE Validation sum*, *MSE training sum* or *Standard sum*. However, a trade-off decision recommends its choice, considering a more straightforward model structure of *DTR (2xGrid+4F)*, and therefore, a lower computational load. *DevdB* took a maximum value of just 0.6 dB in the low-frequency range and up to 3 dB in the high-frequency one.

Regarding the *probe useful bandwidth*, it was accurately predicted. On the other hand, the resonance area presented more disagreement in predictions due to the high variability, which is inherent to the phenomena. Nevertheless, tolerance values are within a typical calibration sensor chart. Thus, the final proposed model provides trustful and fair predictions of the vibration frequency response.

Further research in other *ML* techniques will enrich this study. The standard deviations of the scores obtained in *CV* were high regarding average validation metrics. This inconvenience might be overcome using *ensemble ML* methods like *Random Forest*, where many *DTR* predictors are averaged and might obtain better results. A more detailed survey of *ANN* could also be interesting to assess its potential improvement from current results.

**Roberto San Millán-Castillo** received the B.Sc. degree in telecommunication engineering, minor in Sound & Image from Politécnica University of Madrid, Madrid, Spain, in 2000, the M.Sc. degree in Project management from Nebrija University, Madrid, Spain, in 2012 and the M.Sc. degree in Acoustics Engineering (Research training) from Politécnica University of Madrid, Madrid, Spain, in 2013. He is a Lecturer and researcher and currently pursuing a PhD degree with the Department of Signal Theory and Communications, Rey Juan Carlos University. His research interests include signal processing, machine and statistical learning for applied Acoustics.

**Eduardo Morgado** received the degree in telecommunication engineering from the Carlos III University of Madrid, Leganés, Spain, in 2004, and the PhD degree in telecommunications engineering from Rey Juan Carlos University, Fuenlabrada, Spain, in 2009. He is currently an Associate Professor with the Department of Signal Theory and Communications, Rey Juan Carlos University. His research interests include signal processing for wireless communications with applications to ad hoc, sensor networks, biomedical engineering, and Acoustics.

**Rebeca Goya-Esteban** received the B.Sc. degree in telecommunication engineering from Carlos III University, Madrid, Spain, in 2006, the M.Sc. degree in biomedical engineering from the University of Porto, Porto, Portugal, in 2008, and the PhD degree in telecommunications engineering from Rey Juan Carlos University, Madrid, in 2014. She is currently a Lecturer and Researcher at Rey Juan Carlos University. Her main research interests include time-series analysis, cardiac signal processing, statistical learning, and Acoustics.